\documentclass[aps,floats,twocolumn,showpacs,nofootinbib]{revtex4}

\usepackage{pifont,colortbl,multirow,graphicx,mathrsfs,makeidx,color,epsfig,
fancyhdr,fancybox,calc,amsmath,amsfonts,amssymb,amsthm,latexsym}

\usepackage{epsfig}
\usepackage{epstopdf}
\usepackage{amsmath}
\usepackage[latin1]{inputenc}
\usepackage{graphicx}

\newcommand{\nc}{\newcommand}
\nc{\tcb}{} \nc{\tcr}{}
\nc{\be}{\begin{equation}} \nc{\ee}{\end{equation}}
\nc{\bea}{\begin{eqnarray}} \nc{\eea}{\end{eqnarray}}
\nc{\rds}{{\rm d}s} \nc{\rdt}{{\rm d}t} \nc{\rdr}{{\rm d}r}
\nc{\rdO}{{\rm d}\Omega} \nc{\s}{{\rm S}} \nc{\Pl}{{\rm Planck}}
\nc{\dis}{\displaystyle} \nc{\crit}{_{\rm cr}} \nc{\rd}{{\rm d}}
\nc{\munu}{{\mu\nu}} \nc{\erm}{{\rm e}}

\begin{document}

\title{Black hole dynamical evolution in a Lorentz-violating spacetime}

\author{S. Esposito}
\email{sesposit@na.infn.it}%
\affiliation{Istituto Nazionale di Fisica
Nucleare, Sezione di Napoli, Complesso Universitario di Monte
S.\,Angelo, via Cinthia, I-80126 Naples, Italy}
\author{G. Salesi}
\email{salesi@unibg.it} %
\affiliation{\mbox{Universit\`a Statale di Bergamo, Facolt\`a di
Ingegneria} \mbox{viale Marconi 5, I-24044 Dalmine (BG), Italy}
\mbox{Istituto Nazionale di Fisica Nucleare, Sezione di Milano}
\mbox{via Celoria 16, I-20133 Milan, Italy}}

\

\

\begin{abstract}
\noindent We consider the black hole dynamical evolution in the
framework of a Lorentz-violating spacetime endowed with a
Schwarzchild-like momentum-dependent metric. Large deviations from
the Hawking-Bekenstein predictions are obtained, depending on the
values of the Lorentz-violating parameter $\lambda$ introduced. A
non-trivial evolution comes out, following mainly from the
existence of a non-vanishing minimum mass: for large Lorentz
violations, most of the black hole evaporation takes place in the
initial stage, which is then followed by a stationary stage (whose
duration depends on the value of $\lambda$) where the mass does
not change appreciably. Furthermore, for the final stage of
evolution, our model predicts a sweet slow death of the black
hole, whose ``slowness'' again depends on $\lambda$, in contrast
with the violent final explosion predicted by the standard theory.

\pacs{11.55.Fv; 04.70.Bw; 04.70.Dy; 97.60.-s}
%04.70.-s Physics of black holes (see also 97.60.Lf-in astronomy)
%04.70.Bw Classical black holes
%04.70.Dy Quantum aspects of black holes, evaporation, thermodynamics
%97.60.-s Late stages of stellar evolution (including black holes)
%97.60.Lf Black holes
%11.55.Fv Dispersion relations
%11.30.Cp Lorentz and Poincar\'e invariance

\end{abstract}

\maketitle

\section{Introduction}
\label{sect1}

\noindent Black holes (BH's) are very peculiar physical systems
where, for the enormous forces and the extreme conditions acting
there, a mere classical picture is not sufficient to describe them
and we need taking into account also important quantum effects.
Furthermore, the mechanical description based on general
relativity must be joined with a thermodynamical treatment,
because of the continuous interactions and energy exchanges with
the surrounding environment. Actually, BH's represent a proper
natural laboratory for the investigation of the new edges of
physics. This is particularly true for primordial and for
microscopic (``mini'' or ``micro'') BH's, rather than for stellar
or galactic supermassive ones, because of the extreme smallness of
the Planck scale at which classical and quantum approaches appear
to carry quite different predictions. The most familiar and
relevant quantum effect in BH physics is the so-called
``evaporation'', predicted in the celebrated Hawking-Bekenstein
theory (HB) \cite{HB}, namely the spontaneous radiative particle
emission with progressive mass loss 
and temperature raising, up to a final explosion without any
remnant of the pre-existing BH. This evaporation is not expected
in general relativity, since it is due to a quantum fluctuation on
the event horizon which breaks virtual particle-antiparticle
pairs, releasing the positive energy particles towards the
external surroundings, and confining the negative energy
antiparticles into the interior of the BH, all that resulting in
an effective decreasing of the BH mass.

The observation of the final stages of the BH evaporation \cite{PH},
besides to check general relativity and quantum thermodynamics in extreme
and exotic physical conditions and, above all, the HB theory
itself, could help to find a solution to various important open
problems as, e.g.: the dark matter problem, the cosmological
domain wall problem and the cosmological monopole problem
\cite{Stoj}. Furthermore, the primordial BH density estimated from
observations can severely constraint the matter and energy density 
fluctuations in the early universe.

Let us also remark that the final BH vanishing predicted by the HB
theory is probably a non-physical extrapolation of classical
results to highly non-classical regions. Actually, notwithstanding
general relativity does not require a lower bound to the BH mass,
it is commonly accepted that, at least in ordinary
three-dimensional gravity, the minimum BH mass is of the order of
the Planck mass
 \be M_\Pl = \sqrt{\hbar c^5/G} \sim 10^{19}\,{\rm
GeV} ,
 \ee
which corresponds to an event horizon with a radius of the order
of the Planck length, namely about $10^{-33}$\,cm, and to a
Hawking temperature of the order of $10^{32}$\,K implying a
thermal energy comparable to the BH mass itself. In such
conditions, the emission for Hawking evaporation of only one
photon would cause the vanishing of the whole BH, from which it
follows that a proper thermodynamical description could not be
meaningful.

Besides the unphysical loss rate divergence, the total evaporation
predicted by the standard theory entails other serious problems
and inconsistencies, as the baryon and lepton number
non-conservation, the ``information paradox'', and the
microscopical origin of the entropy \cite{WALD,FN}, which are
essentially due to the complete evaporation of the initial BH.

In the present paper we explore some new predictions of a physical
model introduced recently by one of the Authors (G.S.) \cite{BHE1}, where 
it has been shown that some of the problems mentioned above seem 
to be overcome by adopting a ``deformed'' relativity framework. There, the
energy-momentum dispersion law is Lorentz-violating (LV) and the
Schwarzchild-like metric is momentum-dependent with a Planckian
cut-off. In such a way, net deviations of the basic
thermodynamical quantities from the HB predictions have been
obtained. As a matter of fact, the BH evaporation is expected to quit
at a nonzero critical mass value of the order of the Planck mass,
leaving a zero temperature remnant, and all the semiclassical
corrections to the BH temperature, entropy, and heat capacity are
divergence-free. Then, after reviewing in the following section
the main results of the model proposed, in Sect. \ref{sect3} we
evaluate the modified BH evolution dynamics obtaining highly
non-trivial results which, in the final section, will be analyzed 
and compared with the available experimental data.

\section{Black hole evaporation in a momentum-dependent Schwarzchild metric}
\label{sect2}

\noindent In the last decades high energy (usually planckian)
Lorentz violations have been proposed in many different
experimental and theoretical frameworks as, e.g., (see \cite{BHE1}
and references
therein) GUT's, causal dynamical triangulation, ``extensions'' of the Standard
Model incorporating breaking of Lorentz and CPT symmetries, superstring 
theories, quantum gravity, spacetimes endowed with a non-commutative 
geometry. Let us point up 
that most of the above LV theories seem (implicitly or explictly) to suppose an
essentially non-continuous, discrete spacetime where, as expected
from the uncertainty relations, fundamental momentum and mass-energy 
scales naturally arise. 
\tcb{Hereafter, for simplicity, we shall use the term ``violation" of the Lorentz 
symmetry, but in some of the theories just mentioned (e.g., in the so-called 
``Deformed" or ``Doubly" Special Relativity, where deformed 4-rotation 
generators are considered), although special relativity does not hold anymore, 
an underlying extended Lorentz invariance does exist.}

In the abovementioned paper \cite{BHE1}, following the so-called
Gravity's Rainbow (or Doubly General Relativity) approach
\cite{GR,GAC-Deformed} which generalizes $k$-Minkowski Lie-algebra
to noncommutative curved spacetimes, we have assumed a special LV
momentum-dependent metric where, analogously to the phonon motions
in crystal lattice, only at low energies any particle is allowed
to neglect the quantized structure of the underlying vacuum
geometry. Whilst, at very high energies, particles can effectively
feel the discrete-like structure and the quantum properties of the
medium crossed. Actually, a very general momentum-dependent metric
can be written as follows
 \be \rd s^2 = f^{-2}(p)\rdt^2 -
g^{-2}(p)\rd l^2\,. \label{GRmetric}
 \ee
The form factors $f$ and $g$ are expected to be different from
unity only for planckian momenta, if the LV scale is assumed to be
the Planck energy. One of the most important consequences of
(\ref{GRmetric}) is the modification of the ordinary
momentum-energy dispersion law $\,E^2-p^2=m^2$, by means of
additional terms which vanish in the low momentum limit:
 \be
E^2f^2(p)-p^2g^2(p)=m^2\,.  \label{disp}
 \ee
\tcb{The basic consequence of the replacement of the usual metric
with the ``rainbow" metric in (\ref{GRmetric}) is that the
metric runs, i.e. we have a different metric  for each energy scale.
This does not contradict the principle of relativity, since a non-linear
representation for the Lorentz group is adopted. As a key consequence
of Eq.\,(\ref{disp}), the present model belongs to the general framework
of varying speed of light theories and, as noted in \cite{Ellis}, in such a 
case one should be careful in specifying how diffeomorphism invariance 
or lack of it works, so that the property of invariance of the varying 
speed of light under coordinate transformations holds true. However, 
diffeomorphism transformations change the metrics without changing
the ratio between the speeds of photons with different frequencies. For
the interested reader, an explanation of the structure of these theories,
their Einstein equations, and the impact on BH solutions can be found in
Ref. \cite{Mag}.} 
  
Taking for simplicity $f=1$ in Eq.\,(\ref{disp}), we can write down the above equation
as follows 
 \be E^2 = p^2+m^2+p^2F(p/M)\,, \label{s2}
 \ee
where $M$ indicates a (large) mass scale characterizing the
Lorentz violation. By using a series expansion for $F$, under the
assumption that $M$ is a very large quantity, we can consider only
the lowest order nonzero term in the expansion:
 \be E^2 =
p^2+m^2+\alpha p^2\left(\frac{p}{M}\right)^n
 \ee
($\alpha$ is a dimensionless parameter of order unity). The basic
dispersion law, the most recurring in literature, is the one
corresponding to the lowest exponent, i.e. $n=1$ (see e.g.
\cite{LVN,BALV}, and refs. therein):
 \be E^2 = p^2+m^2+\alpha\frac{p^3}{M}\,. \label{cubic}
 \ee
In ref.\,\cite{BHE1} we adopted a very simple LV metric, namely
 \be f^2(E) = 1, \qquad\qquad\quad g^2(p) = 1-\lambda p\,,
\label{lapses}
 \ee
where the positive parameter $\lambda \sim M_\Pl^{-1}$. This
choice is equivalent to the above first order (n=1) LV dispersion
law (\ref{cubic}) with a negative $\alpha$:
 \be E^2 = p^2 + m^2 -\lambda p^3\,. \label{chosen}
 \ee
The dispersion law (\ref{chosen}) was also previously introduced
in a paper of ours \cite{BALV}, in order to give a simple
explanation for the baryon asymmetry in the Universe. Just because
of the negative sign of the LV term, we succeeded to propose a
straightforward mechanism for generating the observed
matter-antimatter asymmetry through a Lorentz-breakdown energy
scale of the order of the Greisen-Zatsepin-Kuzmin cut-off.

Let us stress that in current Gravity's Rainbow applications to
early universe and to BH's the chosen form factors $f$ and $g$ do
not entail a Planck cut-off and a maximum momentum. By contrast,
in our metric, because of the negative term $-\lambda p$, the
energy vanishes when $p=p_{\rm max}=\dis\frac{1}{\lambda}\sim
M_\Pl$, which plays the role of a ``maximal momentum''
corresponding to the noncontinuous ``granular'' nature of space.
Actually, because the momentum is upperly bounded by $1/\lambda$,
the Heisenberg relations forbid to consider as physically
meaningful any spatial scale smaller than $\lambda$, this
corresponding to an effective ``minimal'' distance between two
distinct space points, a sort of ``step'' of a spatial lattice
that constitutes the vacuum which, at low energy, appears as 
really continuous. \tcb{Notice that such a property directly comes from the fact 
that Eq.\,(\ref{chosen}), differently from what happens in other
models present in the literature, is {\it not} just the leading 
approximation in a series expansion in the small parameter 
$\lambda$ but, rather, Eqs. (\ref{lapses}) are assumed to be the
exact form of the metric, such a form being adopted just
for the related presence of a momentum cutoff. Obviously,
other forms of the metric leading to this property are possible,
but it is remarkable that the predictions obtained through our
approach are not limited to a particular choice, but rather they seem the
physical manifestation of a general result coming from the
presence of a non-vanishing momentum cutoff. In particular, 
it is of some interest the fact that apparently model-independent 
features\footnote{\tcb{P. Nicolini, private communication.}} of 
BH appears even in quite different approaches to quantum gravity 
\cite{NicoliniCasadio}.}

\begin{figure}
\begin{center}
\epsfig{height=5truecm,file=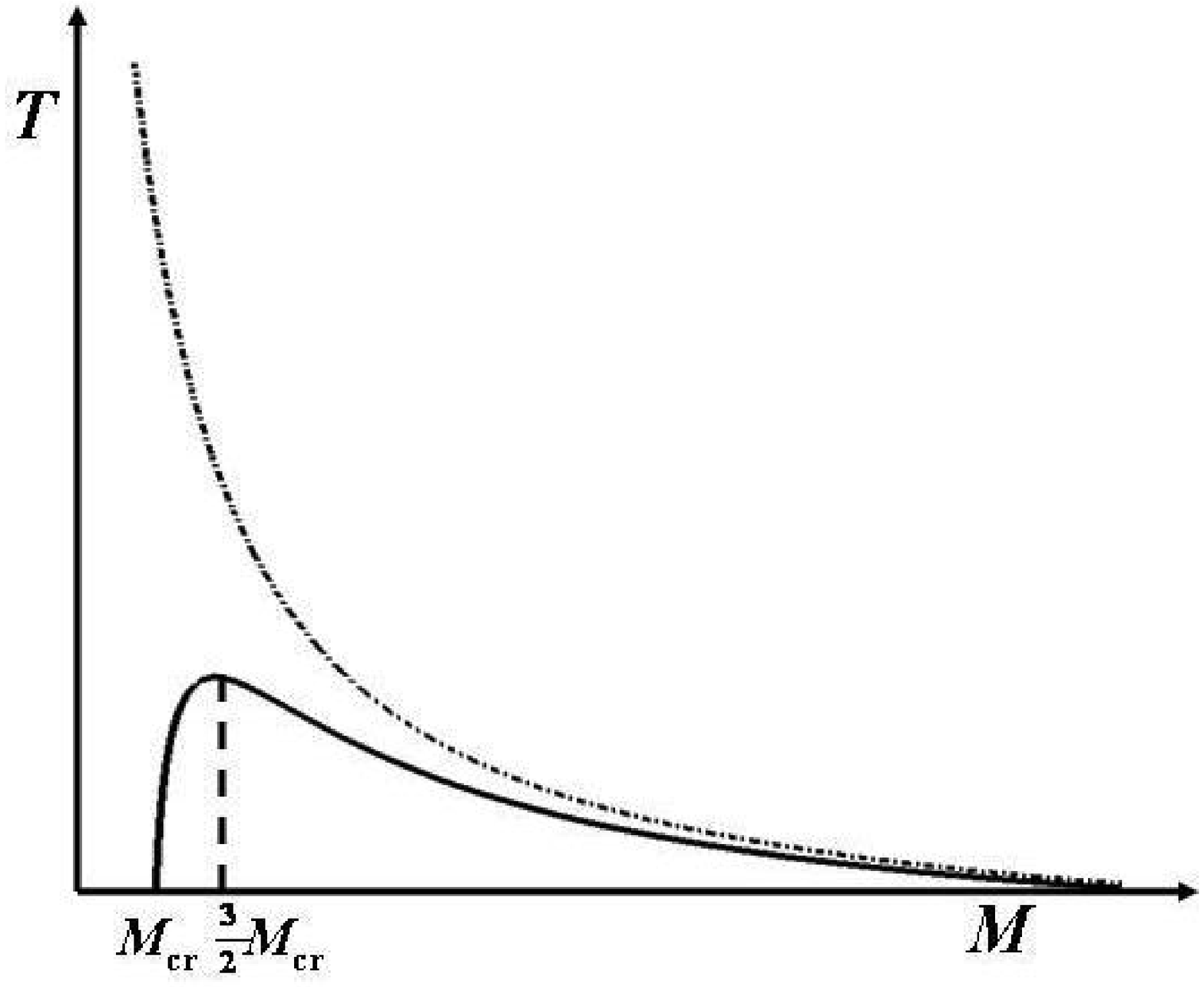} $~~~$
\epsfig{height=5truecm,file=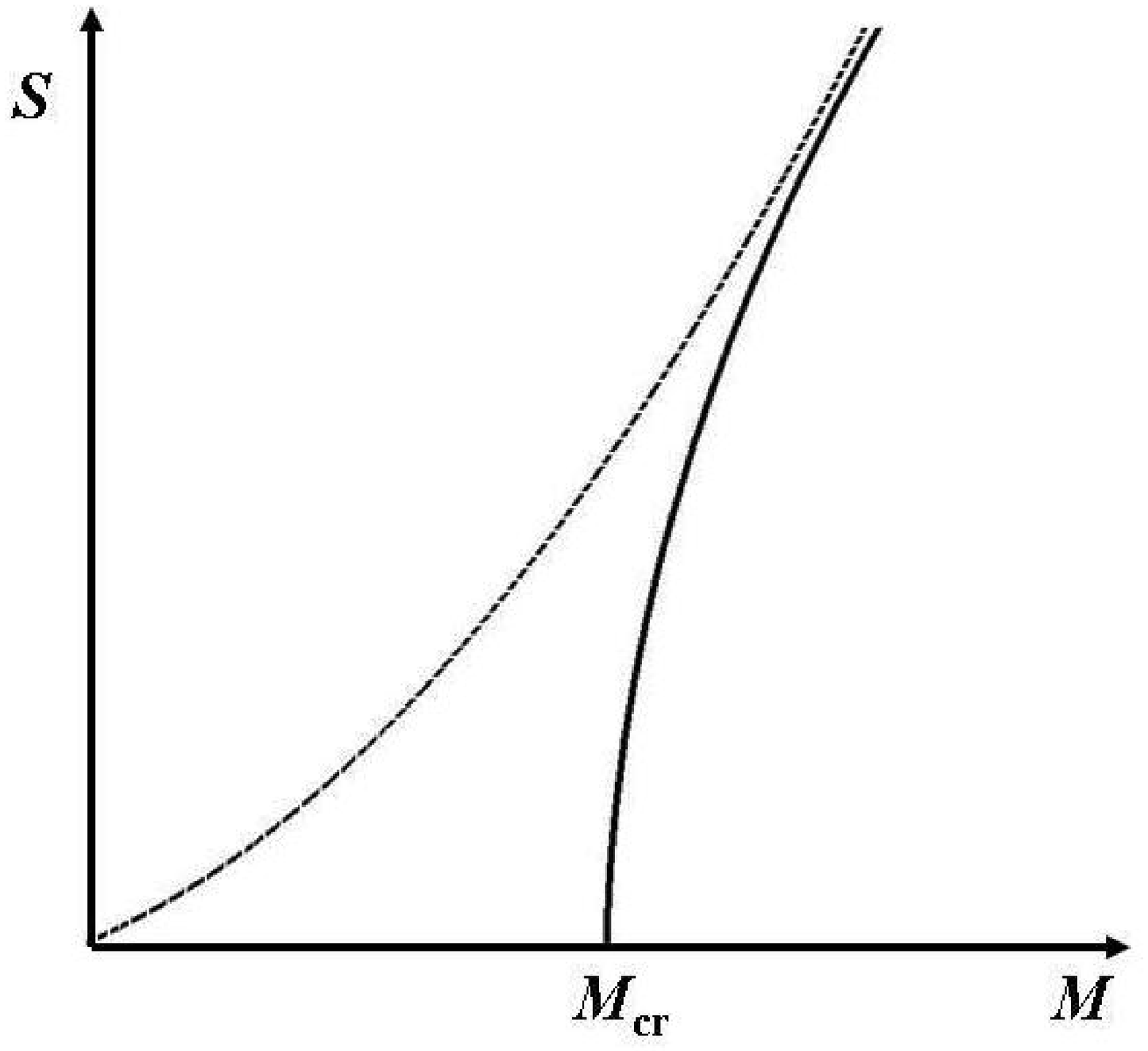}  %
\caption{BH temperature $T$ and entropy $S$ as a function of 
mass $M$, compared with the standard HB plot (dashed lines).}
\label{fig1}
\end{center}
\end{figure}

By adopting the form factors given by (\ref{lapses}) in the
Schwarzchild metric
\bea 
\rds^2 &=& - \left(1-\frac{2GM}{r}\right)\rdt^2 \nonumber \\
& & +\frac{1}{1-\lambda p}\left[\left(1-\frac{2GM}{r}\right)^{-1}\!\!\!\!\rdr^2+ r^2\rdO^2\right] ,
\eea 
in ref.\,\cite{BHE1} we obtained an important correction to the
standard inverse dependence of the BH temperature with respect to
the mass. \tcb{Strictly speaking, in our model particles with different $p$ move in different 
metrics so that, in general, a ``unique" equilibrium temperature is not well-defined. 
However, as in ref.\,\cite{BHE1}, we assume an average behavior for any particle 
described by a unique average temperature for the system since, as we shall see below, 
no large fluctuations are expected in our model. Nevertheless, the BH mass-temperature 
relation shows relevant corrections with respect to the standard case}. Indeed, as a matter of fact, by introducing the minimal LV mass
scale (``critical mass'')
 \be \label{3.11}
 M\crit \equiv \dis\frac{\lambda \hbar c^2}{8\pi G}\,,
 \ee
when $M$ approaches $M\crit$ the temperature sharply deviates from
the HB law \ {($T_{(\lambda=0)} =\hbar c^3/8\pi kGM$)}: 
 \be \label{3.7}
 T = \frac{\hbar c^3}{8\pi kGM} \, \sqrt{1-\frac{M\crit}{M}}
\,.
 \ee
\tcb{Note that for masses larger than the critical value (or the Planck mass), the statistical fluctuations around the average value considered above are obviously very small, while they are non negligible for masses of the order of the critical value. However for such values, due to the Heisenberg's uncertainty relation, the photon momentum is constrained to be almost equal to the extremal cutoff value $p_{\rm max}$, since lower momenta are not allowed when the size of the evaporating BH is of the order of the Planck length. In this regime, the space structure is essentially granular, due to the presence of the cutoff $p_{\rm max}$, which forbids photon momenta greater than (about) the Planck value. Now, if the photons have, practically, the same momentum and then experience the same metric, then the BH temperature is substantially the same for any photon, as given by Eq.\,(\ref{3.7}).\footnote{\tcb{Note, instead, that a unique value of the critical mass is always present, since it is not a dynamical variable of the particles involved but, rather, a parameter of our model.}}
} \\
Owing to evaporation, the BH temperature 
does not diverge at $M=0$ anymore, but shows a finite maximum at
$\dis M=\frac{3}{2}M\crit$ in the latest moments of its life, and
does vanish in the final instant of the evaporation process, at
$M=M\crit\neq 0$ (see Fig.\,\ref{fig1}).
Actually, the most important consequence of our momentum-dependent
metric is that, due to the existence of a maximum momentum, it
exists a minimum mass $M\crit$ for a BH, which therefore ends its
life as a zero-temperature ``extremal'' BH with mass $M\crit$, in
such a way avoiding any singularity formation and also overcoming
all the problems due to the total mass evaporation.\footnote{The
possibility may be considered that the cold BH remnants would be
observed as WIMPs, often considered as the main components of the
dark matter.}

\noindent In the presence of Lorentz violations also the entropy
shows strong deviations from the HB result $S=4\pi GM^2$ when the
mass approaches the critical value $M\crit$ (Fig.\,\ref{fig1}):
 \be
S \simeq \frac{4\pi kGM\crit^{2}}{\hbar c}
\sqrt{\frac{M}{M\crit}-1}\,. \label{expentropmcr}
 \ee
Then the entropy reaches its minimum (zero) together with the
mass, for $M=M\crit$.

\begin{figure}
\begin{center}
\epsfig{height=5truecm,file=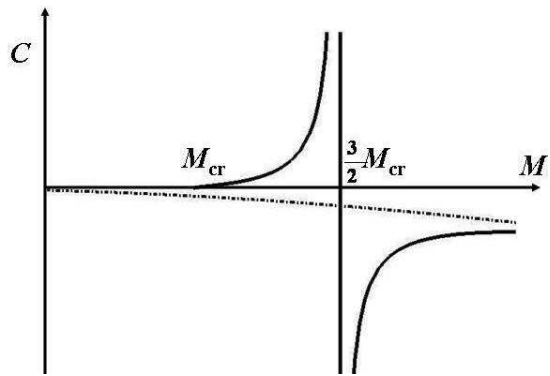} \caption{Heat capacity of a
BH as a function of its mass compared with the standard HB plot
(dashed line).} 
\label{fig3}
\end{center}
\end{figure}

Also another important thermodynamical quantity, the heat
capacity, diverges at $M=\frac{3}{2}M\crit$ (see Fig.\,\ref{fig3}),
corresponding to the maximum BH temperature, and vanishes at the
minimum mass $M\crit$: 
 \be
C \simeq \frac{16 \pi kGM\crit^{2}}{\hbar c} \frac{\displaystyle
\sqrt{\frac{M}{M\crit}-1}}{\displaystyle 3-2\frac{M}{M\crit}}\,.
\ee

\section{Modified Black Hole evolution dynamics}
\label{sect3}

\noindent The momentum-dependent metric adopted here obviously
induces deviations from the time evolution of the BH evaporation as
predicted by the Hawking theory. Here we will investigate just on
such deviations, depending of course on the momentum cutoff
$1/\lambda$.

The thermal radiance $\cal I$ of a photon gas is given by the
product of the light velocity $v$ times the radiation energy
density $\rho$ or, in differential form,
 \be \label{3.1}
\rd {\cal I} = v \, \rd \rho \,.
 \ee
From the dispersion relation (\ref{chosen}) (with $m=0$), the
momentum-dependent group velocity is given by
 \be \label{3.2}
 v = \left|\frac{\rd E}{\rd p}\right| = \frac{|2-3\lambda
p|}{2\sqrt{1-\lambda p}}\,c\,.
 \ee
\tcb{Contrary to the standard case, in our model with a momentum-dependent metric the BH radiation could be, in general, no longer thermal, so that particles in the photon gas might not be distributed in energy according to the equilibrium Bose-Einstein function. Although the correct complete procedure would be that of solving the Boltzmann transport equation for the distribution function in the present case, a simple consideration allows us to avoid such a complication within a reasonable approximation. Indeed, following the discussion after 
Eq.\,(\ref{3.7}), it is likely to expect that the small metric variations induced by small momentum differences around the cutoff  $p_{\rm max}$ are much less effective in destroying  than fast electromagnetic interactions guaranteeing thermodynamical equilibrium. In such a case, we can certainly adopt a thermal distribution for the photons with a rescaled energy $E= c p \sqrt{1-\lambda p}$, so that the energy density is given by:}
 \be \label{3.2b}
\rd \rho = \frac{c}{\pi^2 \hbar^3} \, \frac{p^3 \sqrt{1- \lambda
p}}{\erm^{\frac{c p}{k T} \sqrt{1- \lambda p}} -1} \, \rd p \,.
 \ee
The thermal radiance can then be written in the following general
form:
 \be \label{3.3}
{\cal I} = \frac{c^2}{2 \pi^2 \hbar^3} \, \int_0^{1/\lambda}
\frac{p^3 |2- 3\lambda p|}{\erm^{\frac{c p}{k T} \sqrt{1- \lambda
p}} -1} \, \rd p \,.
 \ee
The lifetime equation may be obtained by considering that such
radiance gives the energy power per unit area emitted by the BH,
 \be \label{3.4}
 {\cal I} = \frac{1}{A} \, \frac{\rd U}{\rd t} \,.
 \ee
In terms of BH mass\footnote{We have, obviously, $U=M c^2$; the BH
surface area $A$ equals $4 \pi$ times the Schwarzchild radius
squared.} $M$ we have
 \be \label{3.5}
{\cal I} = - \frac{c^6}{16 \pi G^2 M^2} \, \frac{\rd M}{\rd t} \,,
 \ee
and substituting Eq.\,(\ref{3.3}):
 \be \label{3.6}
\frac{\rd M}{\rd t} = - \frac{16 G^2 M^2}{2 \pi \hbar^3 c^4} \,
\int_0^{1/\lambda} \frac{p^3 |2- 3\lambda p|}{\erm^{\frac{c p}{k
T} \sqrt{1- \lambda p}} -1} \, \rd p \,.
 \ee
The modified mass-temperature relation to be used here is given by
Eq.\,(\ref{3.7}) \cite{BHE1}. In the standard case with
$\lambda=0$, Eqs. (\ref{3.6}), (\ref{3.7}) reduce to
 \be \label{3.8}
\frac{\rd M}{\rd t} = - \frac{\hbar c^4}{3840 \pi G^2} \,
\frac{1}{M^2} \,.
 \ee
By solving this equation for a BH with initial ($t=0$) mass $M_0$,
the time evolution is just given by
 \be \label{3.9}
M = M_0 \sqrt[3]{1 - \frac{t}{\tau_{(\lambda=0)}}} \,,
 \ee
where
 \be \label{3.10}
\tau_{(\lambda=0)} = \frac{1280 \pi G}{\hbar c^4} \, M_0^3
 \ee
is the BH evaporation time at which $M(t=\tau_{(\lambda=0)})=0$.

For $\lambda \neq 0$ the things sound different, since the final
BH mass is no longer zero. The minimum BH mass can be simply
derived by requiring the argument in the square root in Eq.\,(\ref{3.7}) 
to be non negative, thus obtaining just the minimum
mass value already reported in Eq.\,(\ref{3.11}). The evaporation
time $\tau$ is then now defined by $M(t=\tau)=M\crit$, and can be
obtained by the solution of Eqs. (\ref{3.6}), (\ref{3.7}) . To
this end, it is useful to replace the variables $T,M$ by the
dimensionless quantities
 \be \label{3.12}
\xi = \frac{c}{\lambda k T} \, , \qquad \qquad w =
\frac{M}{M\crit} \,,
 \ee
in terms of which Eq.\,(\ref{3.6}) becomes
 \be \label{3.13}
\frac{\rd w}{\rd t} = - \, \frac{w^2 \gamma(\xi(w))}{\tau_0} \,.
 \ee
Here $\tau_0 = \pi^2 \hbar^2 c^2 \lambda^3/2G$ is a typical
reference time value, while
 \be \label{3.14}
\gamma(\xi) = \frac{1}{2} \, \int_0^1 \frac{y^3 |2-3y|}{\erm^{\xi
y \sqrt{1-y}}-1} \, \rd y
 \ee
(note that $\xi(w) = w \sqrt{w}/\sqrt{w-1}$). The BH time evolution
may  be obtained by Eq.\,(\ref{3.13}) or, more conveniently, by
evaluating numerically the integral in the equivalent equation
 \be \label{3.15}
\frac{t}{\tau_0} = \int_w^{w_0} \frac{\rd w}{w^2 \gamma(\xi(w))}
\,.
 \ee

\begin{figure}
\begin{center}
\epsfig{height=5truecm,file=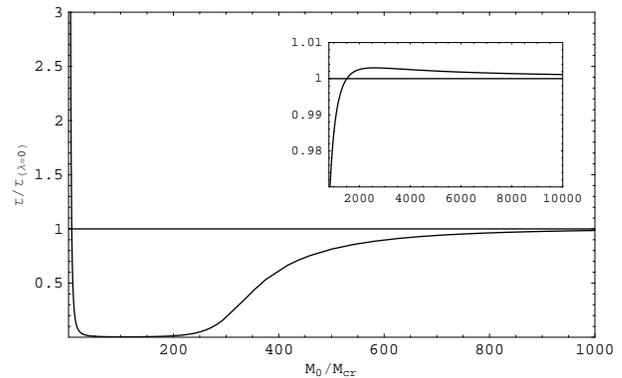} 
\caption{BH lifetime versus initial mass normalized to the
critical value $M\crit$. In the insert we show the peculiar
behavior for very large $M_0/M\crit$: the standard value is
approached asymptotically from the above, and not from below.}
\label{tauriassuntiva}
\end{center}
\end{figure}

The evaporation time $\tau$ for which the BH mass reduces to
$M=M\crit$ is then given by
 \be \label{3.16}
\tau = \frac{\pi^4}{5} \, \frac{1}{w_0^3} \, \tau_{(\lambda =0)}
\int_1^{w_0} \frac{\rd w}{w^2 \gamma(\xi(w))} \,.
 \ee
For small $\lambda$, Eq.\,(\ref{3.16}) is approximated by the
following expression, 
\be \label{3.17}
\tau = \tau_{(\lambda =0)} \left[ 1 + 3 \left( 1 + \frac{270
\zeta(5)}{\pi^4} \right) \frac{M\crit}{M_0} \right] \,,
 \ee
revealing that in such a case the BH lifetime is {\it larger} than
that predicted in the standard theory, as naively expected from
the fact that the final BH mass is now non-vanishing. In general,
however, the situation is more interesting than that, as it is
evident from Fig.\,\ref{tauriassuntiva}. Indeed, we find that for
initial masses $M_0 \lesssim 10\, M\crit$, the evaporation time may
be much greater than the standard value, while for $50\, M\crit
\lesssim M_0 \lesssim 200\, M\crit$ we have $\tau \ll \tau_{(\lambda
=0)}$, reaching a minimum of about $\tau/\tau_{(\lambda =0)} \sim
0.001$. Instead, the standard value of the BH lifetime is
approximately recovered for $M_0 \gtrsim 400\, M\crit$.

\begin{figure}
\begin{center}
\epsfig{height=5truecm,file=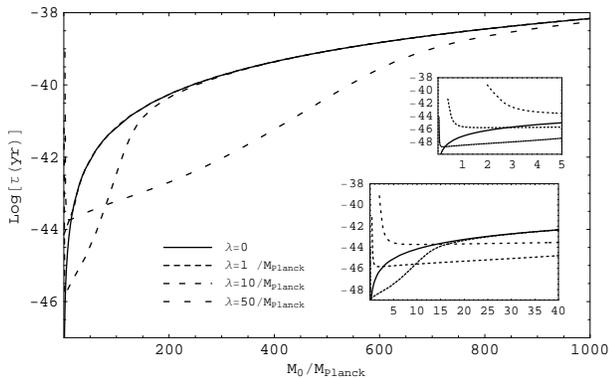} 
\caption{BH lifetime (in absolute units) as function of the
initial mass $M_0$ (in Planck units) for different values of the
Lorentz violating parameter $\lambda$. The endpoints of each curve
for small $M_0$ (see the inserts) correspond to the critical
values $M\crit$ for the considered $\lambda$.} \label{vitamassa}
\end{center}
\end{figure}

\begin{figure}
\begin{center}
\epsfig{height=5truecm,file=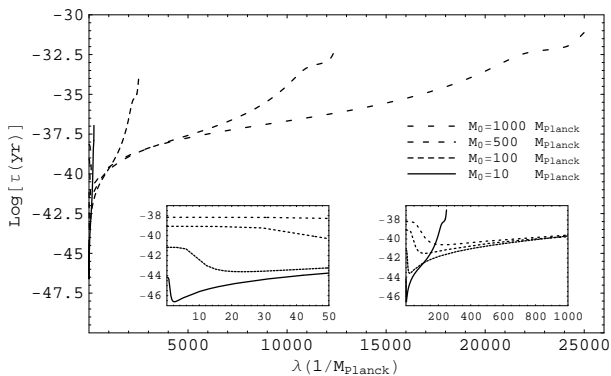} 
\caption{BH lifetime (in absolute units) as function of the
Lorentz violating parameter $\lambda$ (for small values, see the
inserts) for different values of the initial mass $M_0$ (in Planck
units). The endpoints of each curve correspond to the
non-vanishing final value $M\crit$ for the BH mass.}
\label{vitalambda}
\end{center}
\end{figure}

A non-trivial BH evolution dynamics is also evident from the
graphs in Figs. \ref{vitamassa} where the BH lifetime is plotted as a
function of the initial mass (in Planck units) for different
values of the parameter $\lambda$. The sensitivity of the BH dynamics
to this parameter is, instead, deduced from Fig.\,\ref{vitalambda}.
Note that the apparent cutoffs for large $\lambda$ (for the
considered values of $M_0$ are just a manifestation of the
existence of a minimum BH mass, whose value depends on $\lambda$
through Eq.\,(\ref{3.11}). It is then evident that, for large
$\lambda$ (compared to the Planck scale), BH's of mass lower than
the value in Eq.\,(\ref{3.11}) {\it cannot be formed} and,
consequently, no evaporation takes place at all.

\begin{figure}
\begin{center}
\epsfig{height=5truecm,file=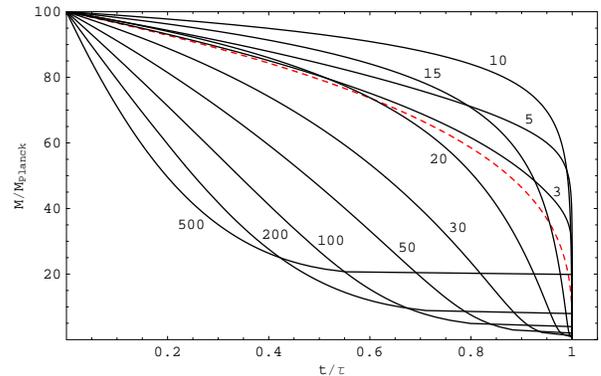} 
\caption{BH mass (in Planck units) versus time (normalized to BH
lifetime) for an initial mass value of $M_0=100\,M_{\rm Planck}$.
The numbers close to each curve gives the value of $\lambda$ in
$M_{\rm Planck}^{-1}$ units. The dashed curve correspond to the
standard case with $\lambda =0$.} \label{masstime}
\end{center}
\end{figure}

The entire dynamics of the BH evaporation may be followed by plotting
BH mass versus time; this is done here in Fig.\,\ref{masstime} for
a typical initial mass value of $M_0=100\,M_{\rm Planck}$.
Obviously, different results are obtained for different values of
the Lorentz-violating parameter $\lambda$ but, as shown in the
mentioned figure, the dependence on $\lambda$ is not monotonic.
For example, for the given value of $M_0$, the mass decrement
before the final stage is slower than in the standard case for
values of $\lambda$ up to approximately $20/M_{\rm Planck}$ (the
curves in Fig.\,\ref{masstime} labeled with $\lambda=3, 5, 10,
15/M_{\rm Planck}$ lie above the one corresponding to the standard
case with $\lambda=0$ for most of the BH life). The reverse is true
for $\lambda$ greater than $20/M_{\rm Planck}$ and, in particular,
for large values of $\lambda$, most of the BH evaporation takes place
in the initial stage, then followed by a stationary stage (lasting
even for more than half of the BH life) where the mass does not change
appreciably.

\begin{figure}
\begin{center}
\epsfig{height=5truecm,file=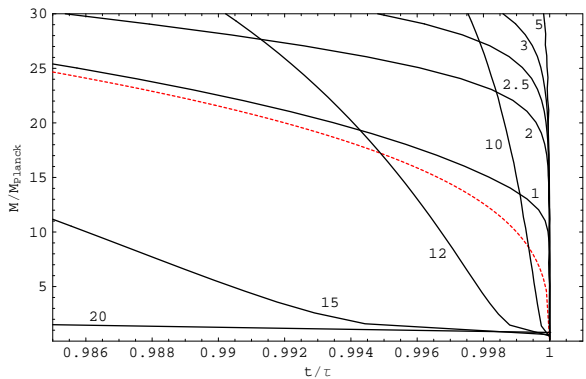} 
\caption{The same as in Fig.\,\ref{masstime} but enlarged near the
final point $t=\tau$. Even for small values of $\lambda$, the
slopes of the curves are smaller than in the standard case (dashed
curve); here this is particularly evident for $\lambda \geq 10$.}
\label{masstimeinsert}
\end{center}
\end{figure}

Furthermore, a very characteristic feature of the BH evolution arises
in our model, regarding the final stage of the life of a BH. In
the standard theory, the BH evolution ends with a final explosion,
characterized by an infinite value of the derivative $\rd M/\rd
t$; that is, the curve $M(t)$ for $\lambda=0$ has a vertical
tangent for $t=\tau$. This never occurs for $\lambda \neq 0$. In
order to show this, we have enlarged the plot in Fig.\,\ref{masstime} 
around $t=\tau$, and the result is shown in Fig.\,\ref{masstimeinsert}. 
It is evident that, even for small values of $\lambda$, the 
slope of the $M(t)$ curve is always smaller than
that in the standard case when approaching the limiting value
$t=\tau$. For very small $\lambda$, the stationary stage discussed
above is increasingly shorter, then recovering the standard result
for $\lambda \rightarrow 0$. Thus our model predicts a BH sweet
slow death without any final explosion, the ``slowness'' of this
stage depending on the value of the Lorentz violating parameter
$\lambda$.

\

\section{Conclusions}

\noindent Starting from our previous introductory work on the subject, we have studied the time evolution of the BH evaporation in a Lorentz-violating spacetime endowed with a Schwarzchild-type momentum-dependent metric. We have computed the BH mean lifetime obtaining net deviations from the HB predictions. Actually, the BH evaporation dynamics results to be ruled by a non-vanishing minimum mass $M\crit$ which is a function of the Lorentz-violating parameter $\lambda$.
We have indeed found that, independently of the strength of Lorentz, for an initial mass of the order of the Planck mass, the BH lifetime can result very larger than the HB prediction $\tau_{(\lambda=0)}$; while, for $M_0$ of the order of 100 $M\crit$, the lifetime results $10^3$ times smaller than $\tau_{(\lambda=0)}$. The standard prediction is recovered only for $M_0 \gtrsim 400\,M\crit$.

Furthermore, we have shown that for large Lorentz violations most of the evaporation takes place at the beginning; then a quasi-stationary era follows where the mass decreases very slowly.

Actually, an important result achieved in the present work regards the final instants
of the BH life. We have indeed found that toward the lifetime end, it is always present an evaporation damping, which starts earlier or later depending on the magnitude of $\lambda$. Consequently, by contrast with Hawking's predictions, our theory does entail a slow death of terminal black holes instead of an infinitely fast evaporation, resulting in a dramatic final gamma-ray burst. This might affect the experimental observation of the Hawking radiation coming out from very light BH's as the so-called ``mini-BH's".

Mini-BH's are supposed to have been created by primordial
density fluctuations in the early universe and, according to the
HB theory, undergo a sufficiently intense evaporation to be
detected by means of suitable experimental devices. Actually,
because of evaporation, primordial BH's with initial mass of the
order of 10$^{15}\,$g (the mass of a typical asteroid) and with
initial Hawking temperature of about 10$^{11}$K would explosively 
vanish just today (while lighter BH's would be already totally 
evaporated). At variance, BH's endowed with a mass larger than 
$M_\odot$, as the ones produced during the gravitational collapse 
of a star, according to HB theory have a lifetime enormously larger 
than $10^{10}$ years, i.e. larger than the estimated age of our 
Universe. On the other hand, supermassive BH's, as galactic ones 
or BH's emerging from star clusters collapse, endowed with a mass 
equal or larger than 10$^6\,M_\odot$, have a temperature still 
smaller with respect to the actual cosmic microwave background 
radiation, so that they are not allowed to evaporate because of 
the second law of thermodynamics. 

Notably, theories which involve additional space dimensions where 
(only) the gravitational force can act ---as, e.g., string theory or
braneworld gravity theories with extra-dimensions 
\cite{KP,RS,Kanti,Kavic,Sendouda,Kawasaki,Dimopoulos}---do carry
significant deviations from the usual BH evaporation dynamics. In fact,
primordial BH's are more easily produced in the Randall-Sundrum 
theoretical framework and, in the presence of extra spatial
dimensions, should undergo a more explosive phase of evaporation.
The increasing of the available number of particle modes for 
the evaporation or some exotic planckian events (as, e.g., phase 
transitions from black string to BH) should result in a strong 
raise of the Hawking radiation luminosity. In some primordial universes
endowed with almost an extra dimension the temperature of primordial 
BH's turns out to be lowered, and the evaporation appears slowing down 
with longer lifetime, so that lighter primordial BH's, not yet evaporated, 
could be detected even in our galaxy.
Furthermore, if LHC is not far above the re-scaled Planck energy (which results lowered to TeV scale in the just mentioned theories), micro-BH's could be abundantly produced in the ATLAS experiment: at the expected luminosities, BH's production rate can be estimated to be of the order of 10 pb, and we could observe several BH's per minute or per second. With a mass of few TeV, the produced BH's would be extremely hot and evaporate almost instantaneously, emitting Hawking radiation composed by particle-antiparticle pairs for all allowed degrees of freedom, at equal rates. 

Instead, according to HB theory, primordial BH's with mass about 10$^{15}\,$g might
be detected by observing high energy gamma rays produced in
the last instants of their lives, which should end with a violent explosion 
equivalent to many teraton hydrogen bombs (typical expected values
of the order of $10^{34}$ erg per $\mu$s).\footnote{Furthermore, it has been suggested that a small BH (of sufficient mass) passing through the Earth would produce a detectable 
acoustic or seismic signal \cite{Seismic}.}
Also with the aim of detecting gamma ray bursts emitted by dying BH's, NASA has recently (2008) 
put in orbit around the Earth the {\em Fermi Gamma-ray Space Telescope} (FGST).
Actually, in our approach we just predict a progressively vanishing evaporation for a terminal mini-BH ending in a cold burnt-out remnant. Therefore, contrary to what abovesaid, because of a very weak terminal Hawking radiation, the actual detection techniques, also by
making recourse to powerful space telescopes, might not be able to observe and measure the weak evaporation of primordial BH's dying nowadays. For analogous reasons, depending on the initial mass and on $\lambda$, the evaporation rate versus time for micro-BH's produced in ultra-high energy particle physics laboratories might involve very weaker signals and very fewer decay products with respect to what currently expected. On the other hand, at the present no such
micro-BH event has been observed at CERN.
In the last fifteen years various high-energy gamma-ray observations have been performed. as e.g. SAS-II \cite{Fichtel75}, COS-B \cite{Bignami}, EGRET \cite{Fichtel83,Fichtel94}. As a matter of fact, in those investigations as well as in the last data from LAT e GBM on FGST, no definitive experimental evidence of terminal BH evaporation via Hawking radiation has been found \cite{Fichtel94,LAT,Bouvier}.
Apart some recent re-interpretations of very short gamma-ray bursts in outdated experiments as KONUS \cite{KONUS}, the current studies present in literature do not strictly relate high energy astrophysical events to strong Hawking radiation pulses, but only upperly constraint the apparent density of terminal mini-BH's. This turns out to be quite consistent with our present predictions on BH time evolution.

Several other interesting cosmological implications deserve the application of our present model; these, however, will be the subject of future studies.

\vspace{0.5cm}

%\noindent {\large{\bf Acknowledgments}}

%\vspace*{0.2cm}

\end{document}